\documentclass[12pt]{article}
\usepackage{amssymb,amsmath,epsfig}

\begin{document}

\title{\bf Dynamics of Magnetized Bulk Viscous Strings in Brans-Dicke Gravity}

\author{M. Sharif \thanks{msharif.math@pu.edu.pk} and Saira Waheed
\thanks{smathematics@hotmail.com}\\
Department of Mathematics, University of the Punjab,\\
Quaid-e-Azam Campus, Lahore-54590, Pakistan.}

\date{}

\maketitle
\begin{abstract}
We explore locally rotationally symmetric Bianchi I universe in
Brans-Dicke gravity with self-interacting potential by using charged
viscous cosmological string fluid. We use a relationship between the
shear and expansion scalars and also take the power law for scalar
field as well as self-interacting potential. It is found that the
resulting universe model maintains its anisotropic nature at all
times due to the proportionality relationship between expansion and
shear scalars. The physical implications of this model are discussed
by using different parameters and their graphs. We conclude that
this model corresponds to an accelerated expanding universe for
particular values of the parameters.
\end{abstract} {\bf
Keywords:} Brans-Dicke theory; Scalar field; Cosmic
expansion.

\section{Introduction}

Current observational measurements obtained from many astronomical
experiments (like Supernova (Ia), WMAP, SDSS, galactic cluster
emission of X-rays, large scale structure and weak lensing etc.)
strengthened the picture of cosmic expansion at an accelerating
rate \cite{1}-\cite{4}. This significant phenomenon of cosmic
expansion is prompted by dark energy (DE), a mysterious unusual
kind of matter containing negative pressure. This is inconsistent
with the strong energy condition and plays a dominant role in the
composition of our universe \cite{5}. The investigation of its
obscure nature is one of the most fascinating issues in modern
cosmology. Consequently, enormous DE proposals including Chaplygin
gas, quintom, k-essence, phantom, quintessence, cosmological
constant etc. have been suggested \cite{6,7}. However, none of
them provides an unambiguous solution to this problem and thus
leaving it as a mystery for the researchers.

General Relativity, in spite of its success in many ways, remained
unsuccessful for describing the reality of DE and some other
cosmological issues. This suggested the exploration of alternative
theories of gravity by taking some modifications in Einstein
Hilbert action \cite{8,9}. For this purpose, numerous modified
gravity theories like Gauss-Bonnet theory, scalar tensor theories,
$f(R)$ and $f(T)$ gravities and recently, $f(R,T)$ gravity theory
etc. have been constructed \cite{10,11}. Among these modified
theories, scalar-tensor theories are the most viable and
interesting candidates of DE. In scalar tensor theories, the
gravity effects are discussed by a tensor as well as a scalar
field \cite{12}-\cite{14}. There are different scalar tensor
theories available in literature, including the scalar tensor
theories formulated by Brans and Dicke, Lyra, Nordtvedt and
Wagoner, Saez and Ballester which are of particular interest
\cite{12,13},\cite{15}-\cite{18}.

Brans-Dicke (BD) gravitational theory is the most prominent and
prevailing case of scalar-tensor theories which provides
convenient solutions to many cosmological issues such as universe
inflation and its late time behavior, coincidence problem, cosmic
acceleration etc \cite{19,20}. Dynamical gravitational constant
$(G=\frac{1}{\phi})$, non-minimal interaction of scalar field with
geometry, compatibility with the Dirac's large number and Mach's
hypotheses as well as weak equivalence principle are some major
facets of this theory \cite{12}-\cite{14}. Various versions of BD
theory are available in literature like generalized and
chameleonic BD theory etc. for different cosmological
implications. Brans-Dicke formalism attracted many researchers to
explore exact cosmological universe models for cosmic expansion
\cite{21}-\cite{23}. Recently, we have investigated exact solution
of the BD field equations by using perfect, anisotropic and
magnetized anisotropic fluids as matter contents \cite{24}.

According to grand unified theories, the phase transition (caused
by the reduction in the temperature below some critical
temperature in the initial epochs of the universe evolution)
results in the creation of some topological defects including
domain walls and cosmic strings etc. These defects are responsible
for density fluctuations and hence lead to a precise description
of structure formation \cite{25}. Since the cosmic strings
interact with gravity, therefore it would be worthwhile to canvas
the string's gravitational and astrophysical consequences. The
study of magnetic field effects in the matter distribution is of
considerable interest as it provides an effective way to
understand the initial phases of cosmic evolution. Bulk viscous
effects in the fluid lead to negative energy field and hence have
a significant impact on the dynamics of the universe \cite{26,27}.
Thus the study of magnetized cosmic stings under the influence of
bulk viscosity leads to a better understanding of the dynamics of
the universe. Various string cosmological universe models have
been investigated in general relativity and scalar tensor theories
\cite{28}-\cite{30}.

This paper deals with the exact BD universe model with magnetized
viscous cosmic strings. The paper is designed in the following
layout. Next section provides BD formulation in the presence of
self-interacting potential for Locally Rotationally Symmetric (LRS)
Bianchi type I (BI) universe model and bulk viscous clouds of
strings with electromagnetic effects. In section \textbf{3}, we
construct the universe model by solving the field equations. We
discuss various physical parameters for the universe model. Lastly,
we present a summary of the obtained results.

\section{LRS Bianchi I Model and BD Formulation}

In a simple BD theory, the BD coupling parameter remains as a
constant, while its modified versions can be obtained by
introducing variable BD parameter, i.e., $\omega(\phi)$ and a
self-interacting potential term. In Jordon frame, the action for
self-interacting BD theory \cite{24} is specified by
\begin{equation}\label{1}
S=\int d^{4}x\sqrt{-g}[L_{\phi}+L_{m}].
\end{equation}
Here $L_{m}$ is the matter contribution and $L_{\phi}$ is the
Lagrangian density having BD scalar field $\phi$ as source and is
given by
\begin{equation}\label{1*}
L_{\phi}=\phi
R-\frac{\omega}{\phi}\phi^{,\alpha}\phi_{,\alpha}-U(\phi),
\end{equation}
where $\omega$ is the BD coupling parameter (which is taken to be
constant), $R$ is the Ricci scalar and $U(\phi)$ represents
self-interacting potential. The self-interacting BD equations
obtained by varying the action with respect to scalar and tensor
fields and are given by
\begin{eqnarray}\nonumber
G_{\mu\nu}&=&\frac{1}{\phi}[\phi_{,\mu;\nu}
-g_{\mu\nu}\Box\phi]+\frac{\omega}{\phi^{2}}[\phi_{,\mu}\phi_{,\nu}
-\frac{1}{2}g_{\mu\nu}\phi_{,\alpha}\phi^{,\alpha}]\\\label{2}&-&\frac{U(\phi)}{2\phi}g_{\mu\nu}
+\frac{T_{\mu\nu}}{\phi},\\\label{3}
\Box\phi&=&\frac{T}{3+2\omega}+\frac{1}{3+
2\omega}[\phi\frac{dU(\phi)}{d\phi}-2U(\phi)].
\end{eqnarray}
Equation (\ref{3}) provides the evolution of scalar field. Here
$T$ and $\Box$ represent the trace of the energy-momentum tensor
and the de'Alembertian operator, respectively. This theory leads
to other modified theories when BD coupling constant takes some
particular values \cite{31,32}. In the limit,
$\phi=\phi_{0},~\omega \rightarrow \infty$ and $T\neq0$, the
respective action and hence the field equations of GR could be
recovered \cite{33,34}.

In order to investigate the universe formation and the initial
epochs of cosmic expansion, the study of Bianchi universe models
is of great significance \cite{35,36}. The LRS BI universe model,
as the simplest generalization of FRW universe model, is described
by an anisotropic spacetime exhibiting spatial homogeneity
\cite{37}
\begin{equation}\label{4}
ds^{2}=dt^{2}-A^{2}(t)dx^{2}-B^{2}(t)(dy^{2}+dz^{2}).
\end{equation}
Here the expansion in $x$ direction is measured by the scale
factor $A$, while in $y$ and $z$ directions it is measured by the
scale factor $B$. The corresponding field equations (\ref{2})
become
\begin{eqnarray}\label{5}
&&\frac{2\dot{A}\dot{B}}{AB}+\frac{\dot{B}^{2}}{B^{2}}=-(\frac{\dot{A}}{A}
+2\frac{\dot{B}}{B})\frac{\dot{\phi}}{\phi}+\frac{\omega}{2}
\frac{\dot{\phi}^{2}}{\phi^{2}}+\frac{U(\phi)}{2\phi}+\frac{T_{00}}{\phi},\\\label{6}
&&2\frac{\ddot{B}}{B}+\frac{\dot{B}^{2}}{B^{2}}=-2\frac{\dot{B}}{B}\frac{\dot{\phi}}{\phi}-\frac{\omega}{2}
\frac{\dot{\phi}^{2}}{\phi^{2}}-\frac{\ddot{\phi}}{\phi}
+\frac{U(\phi)}{2\phi}-\frac{T_{11}}{\phi},\\\label{7}
&&\frac{\ddot{B}}{B}+\frac{\ddot{A}}{A}+\frac{\dot{A}\dot{B}}{AB}=-(\frac{\dot{A}}{A}
+\frac{\dot{B}}{B})\frac{\dot{\phi}}{\phi}-\frac{\omega}{2}
\frac{\dot{\phi}^{2}}{\phi^{2}}-\frac{\ddot{\phi}}{\phi}+\frac{U(\phi)}{2\phi}-\frac{T_{22}}{\phi}.
\end{eqnarray}
The BD scalar wave equation yields
\begin{equation}\label{8}
\ddot{\phi}+3(\frac{\dot{A}}{A}+2\frac{\dot{B}}{B})\dot{\phi}=\frac{T}{2\omega+3}
-\frac{(2\dot{\phi}U(\phi)-\phi\dot{U})}{(2\omega+3)\dot{\phi}}.
\end{equation}
It is interesting to mention here that the energy conservation,
$T^{\mu\nu}_{;\mu}=0$, leads to a linearly dependent equation as
it is an outcome of covariant divergence of the BD equations
(\ref{2}) and (\ref{3}). Thus we leave it and take the field
equations (\ref{5})-(\ref{8}) only.

The average scale factor, the mean and directional Hubble parameters
are
\begin{eqnarray}\label{9*}
&&a(t)=(AB^{2})^{1/3},\quad
H(t)=\frac{1}{3}(\frac{\dot{A}}{A}+2\frac{\dot{B}}{B}),\\\label{9}
&&H_{x}=\frac{\dot{A}}{A},\quad H_{y}=H_{z}=\frac{\dot{B}}{B}.
\end{eqnarray}
The anisotropy measure of expansion $\Delta$, volume $V$, expansion
scalar $\Theta$, deceleration parameter $q$ as well as shear scalar
$\sigma$ can be written as
\begin{eqnarray}\label{10}
\Delta&=&\frac{1}{3}\sum^{3}_{i=1}(\frac{H_{i}-H}{H})^{2},\quad
V=a^3(t)=AB^{2},\\\label{11}
\Theta&=&u^{\alpha}_{;\alpha}=\frac{\dot{A}}{A}+2\frac{\dot{B}}{B},\quad
q=\frac{d}{dt}(\frac{1}{H})-1,\quad
\sigma=\frac{1}{\sqrt{3}}(\frac{\dot{A}}{A}-\frac{\dot{B}}{B}).
\end{eqnarray}
Here $\Delta=0$ corresponds to isotropic expansion of the universe
model.

\section{Model for the Magnetized Viscous Cosmic String Fluid}

In this section, we discuss BI model with electromagnetic bulk
viscous cloud of strings as background fluid distribution given by
the energy-momentum tensor \cite{27}
\begin{equation}\label{12}
T^{\nu}_{\mu}=(\rho+P_{eff})u^{\nu}u_{\mu}-P_{eff}\delta^{\nu}_{\mu}-\lambda
x^{\nu}_{\mu}+E^{\nu}_{\mu},
\end{equation}
where $u^{\mu}$ is the particle's four velocity, $\lambda$ is the
string tension density, $E^{\nu}_{\mu}$ denotes the electromagnetic
part of energy-momentum tensor and $x^{\mu}$ is the spacelike unit
vector that provides the string direction. Here we take
$u^{\mu}=(1,0,0,0)$ and $x^{\mu}=(0,A^{-1},0,0)$ which satisfy the
relations
\begin{equation*}
u^{\mu}u_{\mu}=1=-x^{\mu}x_{\mu}, \quad u^{\mu}x_{\mu}=0.
\end{equation*}
The effective pressure $P_{eff}$ is defined as the sum of isotropic
and viscous pressures, i.e., $P_{eff}=P_{I}+P_{vis}$. We consider
here the dust case, i.e., $P_{I}=0$. Moreover, $P_{vis}=-\xi\Theta$,
where $\xi$ denotes the bulk viscosity coefficient. The
electromagnetic part of $T^{\nu}_{\mu}$ is given by
\begin{equation}\label{13}
E^{\nu}_{\mu}=\overline{\mu}[|h|^{2}(u_{\mu}u^{\nu}+1/2\delta^{\nu}_{\mu})-h_{\mu}h^{\nu}],
\end{equation}
where $h_{\mu}$ is the magnetic flux vector
\begin{equation}\label{14}
h_{\mu}=\frac{\sqrt{-g}}{2\overline{\mu}}\epsilon_{\mu\nu\alpha\beta}F^{\alpha\beta}u^{\nu},
\end{equation}
the terms $\overline{\mu},~F^{\alpha\beta}$ and
$\epsilon_{\mu\nu\alpha\beta}$ denote the magnetic permeability, the
electromagnetic field tensor and the Levi-Civita tensor,
respectively. Moreover, $\rho$ represents the proper density of
strings being the sum of the particle density $\rho_{p}$ (as the
particles are attached to these strings) and string tension density
$\lambda$ is specified by $\rho=\rho_{p}+\lambda$.

We assume that the magnetic field is generated in $yz$ plane as
its source is the electric current that flows in $x$ direction.
Here the magnetic flux vector has only one non-zero component
$h_{1}$. Moreover, the assumption of infinitely large conductivity
along with finite current (in magnetohydrodynamics limit) leads to
$F_{10}=F_{20}=F_{30}=0$ \cite{38}. Thus, all the electromagnetic
field tensor components vanish except $F_{23}$. Using Maxwell's
equations
\begin{equation*}
F_{\alpha\beta;\rho}+F_{\beta
\rho;\alpha}+F_{\rho\alpha;\beta}=0,\quad F^{\alpha\beta}_{;\rho}=0,
\end{equation*}
we find $F_{23}=\Pi$, a constant. The matter tensor (\ref{12}) has
the trace $T=\rho+\lambda-3\xi\Theta$. The non-zero component of
magnetic flux vector will be $h_{1}=\frac{\Pi}{\overline{\mu}B}$,
yielding the corresponding components of $E^{\nu}_{\mu}$
\begin{equation}\nonumber
E^{0}_{0}=E^{1}_{1}=\frac{-\Pi^{2}}{2\overline{\mu}B^{4}}=-E^{2}_{2}=-E^{3}_{3}.
\end{equation}
There are seven unknowns namely $\phi,~U,~A,~B,~\rho,~\lambda,~\xi$
and only three independent field equations. For a closed set of
equations, we take the following assumptions:
\begin{itemize}
\item $A=B^{n};~n\neq1$, this condition is constructed by setting the ratio $\frac{\sigma}{\Theta}$
as constant, i.e., we assume the proportional relationship between
the expansion and shear scalar \cite{39}-\cite{42}.
\item $\phi=\phi_{0}B^{\delta}$, a well-known power law relationship which
indicates that the evolution of BD scalar field $\phi$ is
dependent on the scale factor. Moreover, $\delta>0$ for expanding
solutions \cite{24}.
\item $U=U_{0}\phi^{k}$, a power law ansatz for potential in
which $k$ is any non-zero integer which can further be written as
$U=U_{0}B^{k\delta}$.
\item Also, we take $\xi\Theta=M_{1}$, by setting the expansion scalar
inversely proportional to bulk viscosity coefficient \cite{27}
which means that the rate of cosmic expansion decreases as the
viscosity increases. Here $M_{1}$ is any positive constant.
 \end{itemize}
The field equations for the fluid (\ref{12}) turn out to be
\begin{eqnarray}\label{15}
(2n+1)\frac{\dot{B}^{2}}{B^{2}}&=&-(n+2)\frac{\dot{B}}{B}\frac{\dot{\phi}}{\phi}
+\frac{\omega}{2}\frac{\dot{\phi}^{2}}{\phi^{2}}
+\frac{U(\phi)}{2\phi}+\frac{\rho}{\phi}-\frac{\Pi^{2}}{2\overline{\mu}B^{4}\phi},\\\nonumber
2\frac{\ddot{B}}{B}+\frac{\dot{B}^{2}}{B^{2}}&=&-2\frac{\dot{B}}{B}\frac{\dot{\phi}}{\phi}
-\frac{\omega}{2}\frac{\dot{\phi}^{2}}{\phi^{2}}
-\frac{\ddot{\phi}}{\phi}+\frac{U(\phi)}{2\phi}
+\frac{\xi\Theta}{\phi}+\frac{\lambda}{\phi}-\frac{\Pi^{2}}{2\overline{\mu}B^{4}\phi},\\\label{16}\\\nonumber
(n+1)\frac{\ddot{B}}{B}+n^{2}\frac{\dot{B}^{2}}{B^{2}}
&=&-(n+1)\frac{\dot{B}}{B}\frac{\dot{\phi}}{\phi}-\frac{\omega}{2}
\frac{\dot{\phi}^{2}}{\phi^{2}}-\frac{\ddot{\phi}}{\phi}
+\frac{U(\phi)}{2\phi}+\frac{\xi\Theta}{\phi}\\\label{17}
&+&\frac{\Pi^{2}}{2\overline{\mu}B^{4}\phi},\\\label{18}
\ddot{\phi}+(n+2)\frac{\dot{B}}{B}\dot{\phi}&=&\frac{(\rho+\lambda+3\xi\Theta)}{(2\omega+3)}
+\frac{(\phi\dot{U}-2\dot{\phi}U(\phi))}{(2\omega+3)\dot{\phi}},
\end{eqnarray}
where we have used $A=B^{n}$.

Now there are three independent field equations and three unknowns
namely $\rho,~B$ and $\lambda$. Equation (\ref{17}) yields
\begin{eqnarray}\nonumber
2\ddot{B}+\frac{(2n(n+\delta)+(\omega_{0}+2)\delta^{2})\dot{B}^{2}}
{(n+\delta+1)B}&=&\frac{2M_{1}B^{(1-\delta)}}{\phi_{0}(\delta+n+1)}
+\frac{\Pi^{2}B^{-(\delta+3)}}{\overline{\mu}\phi_{0}(\delta+n+1)}\\\nonumber
&+&\frac{U_{0}B^{(\delta k-\delta+1)}}{\phi_{0}(\delta+n+1)}
\end{eqnarray}
whose solution is
\begin{eqnarray}\nonumber
\dot{B}^{2}&=&\frac{2M_{1}B^{(2-\delta)}}{\phi_{0}(\delta+n+1)(2-\delta+s)}
-\frac{\Pi^{2}B^{-(\delta+2)}}{\overline{\mu}\phi_{0}(\delta+n+1)(\delta+2-s)}\\\label{19}
&+&\frac{U_{0}B^{(\delta k+2-\delta)}}{\phi_{0}(\delta+n+1)(\delta
k+2-\delta+s)}+\frac{M_{2}}{B^{s}},
\end{eqnarray}
where $M_{2}$ is a constant of integration (taken to be positive)
and $s$ is a constant given by
\begin{equation*}
s=\frac{2n(n+\delta)+(\omega+2)\delta^{2}}{(n+\delta+1)}.
\end{equation*}
Equation (\ref{19}) leads to
\begin{eqnarray}\nonumber
dt&=&[\frac{2M_{1}B^{(2-\delta)}}{\phi_{0}(\delta+n+1)(2-\delta+s)}
-\frac{\Pi^{2}B^{-(\delta+2)}}{\overline{\mu}\phi_{0}(\delta+n+1)(\delta+2-s)}\\\label{20}
&+&\frac{U_{0}B^{(\delta k+2-\delta)}}{\phi_{0}(\delta+n+1) (\delta
k+2-\delta+s)}+\frac{M_{2}}{B^{s}}]^{-1/2}dB.
\end{eqnarray}

Introducing the notions $B=\tau$ for time and $X,~Y,~Z$ for the
space coordinates, the corresponding BI universe model becomes
\begin{eqnarray}\nonumber
ds^{2}&=&[\frac{2M_{1}\tau^{(2-\delta)}}{\phi_{0}(\delta+n+1)(2-\delta+s)}
-\frac{\Pi^{2}\tau^{-(\delta+2)}}{\overline{\mu}\phi_{0}(\delta+n+1)(\delta+2-s)}\\\nonumber
&+&\frac{U_{0}\tau^{(\delta
k+2-\delta)}}{\phi_{0}(\delta+n+1)(\delta k+2-\alpha+s)}
+\frac{M_{2}}{\tau^{s}}]^{-1}d\tau^{2}-\tau^{2n}dX^{2}\\\label{21}
&-&\tau^{2}(dY^{2}+dZ^{2}).
\end{eqnarray}
Equation (\ref{15}) yields the density as
\begin{equation}\nonumber
\rho(B)=[(1+2n)+\delta(n+2)-\frac{\omega\delta^{2}}{2}]\phi_{0}\dot{B}^{2}B^{(\delta-2)}
+\frac{U_{0}B^{2}}{2}+\frac{\Pi^{2}}{2\overline{\mu}B^{4}}.
\end{equation}
Substituting the value from Eq.(\ref{19}), it follows that
\begin{eqnarray}\nonumber
\rho(\tau)&=&\phi_{0}(1+2n+\delta(n+2)-\frac{\omega\delta^{2}}{2})\frac{M_{2}}{\tau^{s-\delta+2}}
+[(1+2n+\delta(n+2)\\\nonumber
&-&\frac{\omega\delta^{2}}{2})\frac{2M_{1}}{(\delta+n+1)(2-\delta+s)}]
-[\frac{(1+2n+\delta(n+2)-\frac{\omega\delta^{2}}{2})}{(\delta+n+1)(2+\delta-s)}\\\label{22}
&-&\frac{1}{2}]\frac{\Pi^{2}}{\overline{\mu}\tau^{4}}-[\frac{1}{2}
-\frac{(1+2n+\delta(n+2)-\frac{\omega\delta^{2}}{2})}{(\delta+n+1)(\delta
k+2+s-\delta)}]U_{0} \tau^{\delta k}.
\end{eqnarray}
The string tension density $\lambda$ can be expressed from
Eqs.(\ref{16}) and (\ref{18}) as
\begin{eqnarray}\nonumber
\lambda(\tau)&=&\frac{(3+2\omega)\phi_{0}\tau^{\delta}}{(4+2\omega)}[\frac{2\ddot{\tau}}{\tau}+
(1+\frac{\omega\delta^{2}}{2}+n\delta+4\delta)\frac{\dot{\tau}^{2}}{\tau^2}
-\frac{U_{0}\tau^{\delta(k-1)}}{\phi_{0}(3+2\omega)}\\\label{23}&
\times&(1+2\omega+k)-\frac{M_{1}(6+2\omega)}{(3+2\omega)\phi_{0}\tau^{\delta}}
+\frac{\Pi^{2}}{2\overline{\mu}\tau^{\delta+4}\phi_{0}}]-\frac{\rho}{4+2\omega},
\end{eqnarray}
where
\begin{eqnarray}\nonumber
\frac{\ddot{\tau}}{\tau}&=&\frac{1}{(\delta+n+1)}[\frac{M_{1}(2-\delta)\tau^{-\delta}}{\phi_{0}(s+2-\delta)}
-\frac{(\delta+2)\Pi^{2}\tau^{-(\delta+4)}}{2\overline{\mu}\phi_{0}(s-\delta-2)}
\\\nonumber&+&\frac{U_{0}(\delta(k-1)+2)\tau^{\delta(k-1)}}{2\phi_{0}(\delta(k-1)+2+s)}],
\end{eqnarray}
$\rho$ is given by Eq.(\ref{22}) and $\frac{\dot{\tau}^2}{\tau^2}$
can be calculated from Eq.(\ref{19}).

The density of the particles is
\begin{eqnarray}\nonumber
\rho_{p}=\rho-\lambda&=&-\frac{\rho(5+2\omega)}{4+2\omega}
-\frac{(3+2\omega)\phi_{0}\tau^{\delta}}{4+2\omega}[\frac{2\ddot{\tau}}{\tau}+
(1+\frac{\omega\delta^{2}}{2}\\\nonumber&+&n\delta+4\delta)\frac{\dot{\tau}^{2}}{\tau^2}
-\frac{U_{0}\tau^{\delta(k-1)}}{\phi_{0}(3+2\omega)}
(1+2\omega+k)\\\label{24}&-&\frac{M_{1}(6+2\omega)}{(3+2\omega)\phi_{0}\tau^{\delta}}
+\frac{\Pi^{2}}{2\overline{\mu}\tau^{\delta+4}\phi_{0}}].
\end{eqnarray}
The directional Hubble, mean Hubble and deceleration parameters
become
\begin{eqnarray}\nonumber
H_{x}=nH_{y}&=&nH_{z}=n[\frac{2M_{1}\tau^{-\delta}}{\phi_{0}(\delta+n+1)(2-\delta+s)}
-\frac{\Pi^{2}(\delta+n+1)^{-1}}{\overline{\mu}\phi_{0}(\delta+2-s)}\\\nonumber
&\times&\tau^{-(\delta+4)}-\frac{U_{0}\tau^{(k-1)\delta}}{\phi_{0}(\delta+n+1)(k\delta-\delta+s)}
+\frac{M_{2}}{\tau^{(s+2)}}]^{1/2},\\\nonumber
H&=&\frac{(n+2)}{3}[\frac{2M_{1}\tau^{-\delta}}{\phi_{0}(\delta+n+1)(2-\delta+s)}
-\frac{\Pi^{2}(\delta+n+1)^{-1}}{\overline{\mu}\phi_{0}(\delta+2-s)}\\\nonumber
&\times&\tau^{-(\delta+4)}-\frac{U_{0}\tau^{(k-1)\delta}}{\phi_{0}(\delta+n+1)(k\delta-\delta+s)}
+\frac{M_{2}}{\tau^{(s+2)}}]^{1/2},\\\nonumber
q&=&\frac{3(s+1)}{(n+2)}-1-\frac{3\tau^2}{(n+2)(\delta+n+1)\dot{\tau}^2}[\frac{M_{1}}{\phi_{0}\tau^{\delta}}
+\frac{\Pi^{2}\tau^{-(\delta+4)}}{2\overline{\mu}\phi_{0}}\\\nonumber
&+&\frac{U_{0}\tau^{\delta(k-1)}}{2\phi_{0}}].
\end{eqnarray}
The expansion scalar is
\begin{eqnarray}\nonumber
\Theta&=&(n+2)[\frac{2M_{1}\tau^{-\delta}}{\phi_{0}(\delta+n+1)(2-\delta+s)}
-\frac{\Pi^{2}\tau^{-(\delta+4)}}{\overline{\mu}\phi_{0}(\delta+n+1)(\delta+2-s)}
\\\label{25}&+&\frac{U_{0}\tau^{(k-1)\delta}}{\phi_{0}(\delta+n+1)
(k\delta-\delta+s)}+\frac{M_{2}}{\tau^{(s+2)}}]^{1/2}.
\end{eqnarray}
The coefficient of bulk viscosity takes the form
\begin{eqnarray}\nonumber
\xi&=&\frac{M_{1}}{(n+2)}[\frac{2M_{1}\tau^{-\delta}}{\phi_{0}(\delta+n+1)(2-\delta+s)}
-\frac{\Pi^{2}\tau^{-(\delta+4)}}{\overline{\mu}\phi_{0}(\delta+n+1)(\delta+2-s)}
\\\label{26}&+&\frac{U_{0}\tau^{(k-1)\delta}}{\phi_{0}(\delta+n+1)(k-\delta+s)}+\frac{M_{2}}{\tau^{(s+2)}}]^{-1/2}.
\end{eqnarray}
The shear scalar is
\begin{eqnarray}\nonumber
\sigma&=&\frac{(1-n)}{\sqrt{3}}[\frac{2M_{1}\tau^{-\delta}}{\phi_{0}(\delta+n+1)(2-\delta+s)}
-\frac{\Pi^{2}\tau^{-(\delta+4)}}{\overline{\mu}\phi_{0}(\delta+n+1)(\delta+2-s)}\\\label{27}
&-&\frac{U_{0}\tau^{(k-1)\delta}}{\phi_{0}(\delta+n+1)(k\delta-\delta+s)}+\frac{M_{2}}{\tau^{(s+2)}}]^{1/2}.
\end{eqnarray}

The volume for the model takes the form $V=\tau^{n+2}$ which is zero
initially and becomes divergent when $\tau\rightarrow\infty$
indicating that the expansion of the universe starts from zero to
infinite volume. The anisotropic measure of expansion leads to
$\Delta=\frac{2(n-1)^2}{(n+2)^2}$ which is constant (as we have
taken the shear scalar proportional to expansion scalar) and it
becomes zero for $n=1$. In our case, $n\neq1$, thus the universe
model remains anisotropic throughout the cosmic time. The energy
density remains positive for the allowed range of parameters as
shown in Figure \textbf{1(a)}. However, it becomes infinite at
initial epoch but decreases for final phases of the universe
evolution. The energy density approaches to a positive value due to
viscosity, given by
\begin{equation*}
\rho=\frac{2M_{1}}{(\delta+n+1)(s+2-\delta)}[1+2n-\frac{\omega\delta^2}{2}+\delta(n+2)]
\end{equation*}
as $\tau\rightarrow\infty$. If we take viscosity to be negligible,
then the density approaches to zero for later times thus
representing an empty universe in future.

The string tension density $\lambda$ increases with the passage of
time and then remains a constant for future evolution of the
universe as shown in Figure \textbf{1(b)}. The particle density
$\rho_{p}$ exhibits a similar behavior as does the energy density.
It decreases from infinite value (at $\tau=0$) to a certain constant
due to viscosity effects and then remains constant for the future
evolution as shown in Figure \textbf{2(a)}.
\begin{figure}
\centering \epsfig{file=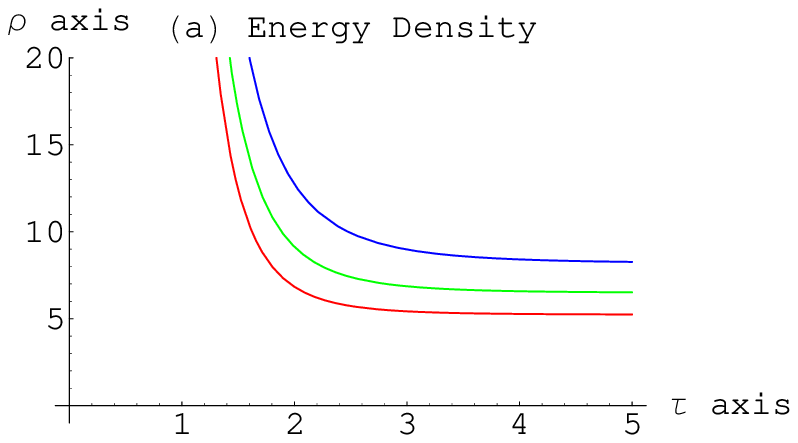,width=.45\linewidth}
\epsfig{file=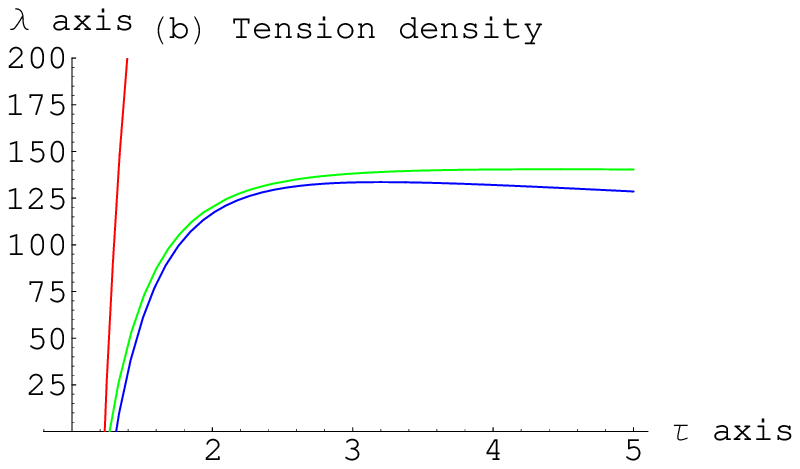,width=.45\linewidth} \caption{Plots (a) and
(b) show the energy density $\rho$ and the string tension density
$\lambda$ versus time $\tau$. Here, red, green and blue lines
correspond to $\omega=-1.7,~-1.8$ and $-1.9$ respectively. Also, we
have taken $\delta=
2,~M_{1}=4,~M_{2}=3,~n=3,~\bar{\mu}=0.5,~k=-4,~\Pi^2=3,~U_{0}=1,$
and $\phi_{0}=1$.}
\end{figure}
The model results in dynamical deceleration parameter and
consequently leads to negative as well as positive values for
certain choices of parameters. For later time with
$k<0,~-2<\omega<-1.5,~0<\delta<1$ and $n>2$, the deceleration
parameter becomes negative, indicating accelerated expanding
behavior of the universe model given by
\begin{equation*}
q=\frac{3(s+1)}{n+2}-1-\frac{3(2-\delta+s)}{(n+2)(\delta+n+1)}.
\end{equation*}
However, for other choices of parameters, it exhibits decelerating
behavior.

The graph of deceleration parameter for particular choices of the
parameters is given in Figure \textbf{2(b)}. It is clear that the
model represents accelerated expanding universe as the deceleration
parameter attains small negative values as shown in this figure.
\begin{figure}
\centering \epsfig{file=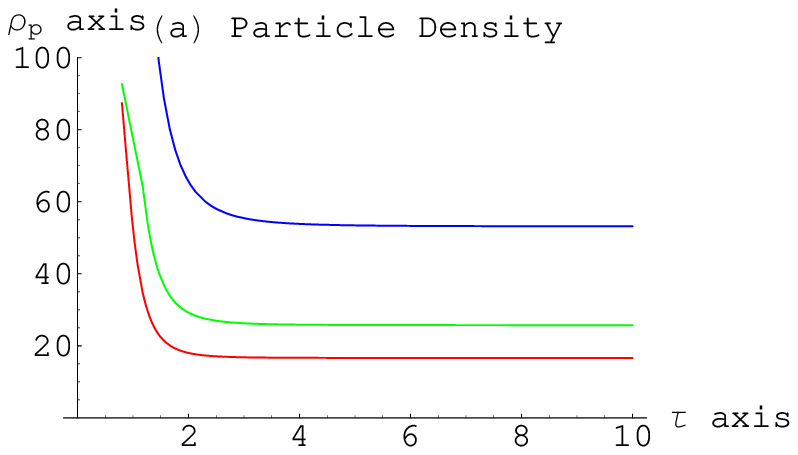,width=.45\linewidth}
\epsfig{file=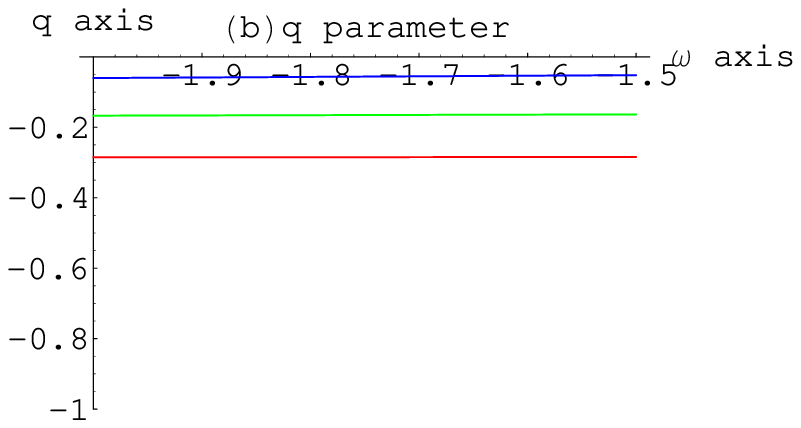,width=.45\linewidth} \caption{Plot (a) shows
the particle density $\rho_{p}$ versus time $\tau$  and (b)
indicates the deceleration parameter $q$ versus BD parameter
$\omega$. In graph (b), we have taken $n=2$, the blue, green and red
lines correspond to $\delta=0.2,~0.4$ and $0.6$ respectively.}
\end{figure}
The graphical illustration for expansion scalar is given in Figure
\textbf{3(a)}, while the shear scalar, directional and mean Hubble
parameters exhibit a similar behavior. These parameters go to zero
when $\tau\rightarrow \infty$ and diverge for initial time
indicating the beginning of the universe model with a big bang
explosion as shown in Figure \textbf{3(a)}. Figure \textbf{3(b)}
indicates the graphical behavior of viscosity parameter $\xi$ which
is opposite to the expansion scalar (in accordance with our
assumption). The viscosity parameter is negligible at the initial
epoch and increases with the passage of time as shown in Figure
\textbf{3(b)}. Hence it prevents the universe to be empty for its
future evolution.
\begin{figure}
\centering \epsfig{file=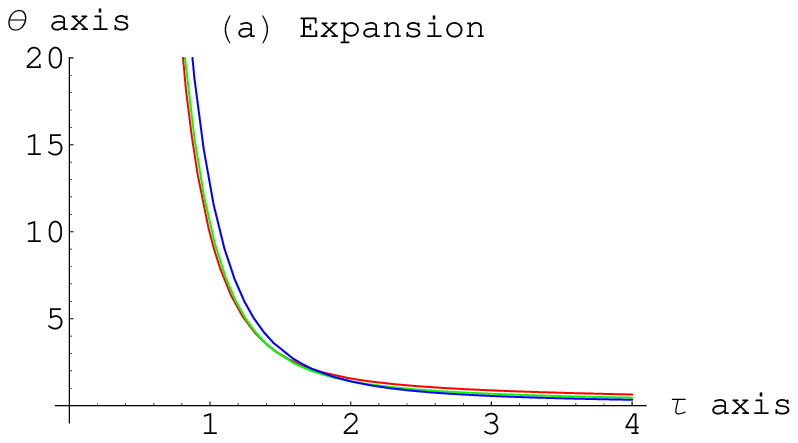,width=.45\linewidth}
\epsfig{file=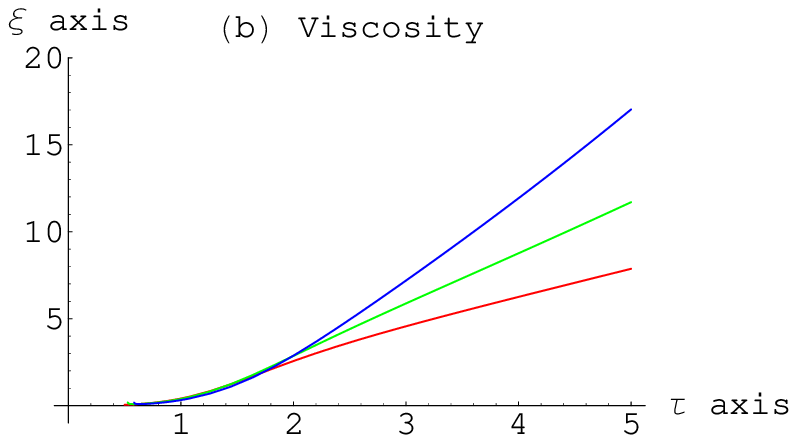,width=.45\linewidth} \caption{Plots (a) and
(b) represent the expansion parameter $\Theta$ and viscosity $\xi$
respectively. Here,
$M_{1}=4,~M_{2}=3,~n=3,~\bar{\mu}=0.5,~k=-4,~\Pi^2=3,~ U_{0}=1,$ and
$\phi_{0}=1$. The red, green and blue colors correspond to
$\omega=-1.7,~-1.8$ and $-1.9$ respectively.}
\end{figure}

\section{Summary}

It is argued that magnetic field has cosmological origin as there
was highly ionized matter that was coupled to magnetic field which
further leads to neutral matter as a consequence of universe
expansion. It is interesting to discuss the magnetic field effects
on the expansion history of the universe. Moreover, bulk viscosity
effects and scalar field have important role for obtaining
accelerated expanding universe model. Thus it would be worthwhile to
discuss BI universe filled with magnetized bulk viscous strings for
the discussion of early and late stages of the universe in scalar
tensor theories. This paper investigates the cosmological model in
self-interacting BD gravity with magnetized bulk viscous cloud of
strings by taking certain physical conditions. All the cosmological
parameters depend upon the values of BD parameter $\omega$,
parameter $n$ as well as parameter $\delta$ (that appear due to
scalar field). We have discussed the resulting model using different
physical parameters through graphs. The results are summarized as
follows.
\begin{itemize}
\item According to Hubble \cite{43}, there would have been an
infinitely hot and dense universe in its early phase. In fact, all
the density of the universe was concentrated at a single point and
hence initially there was zero volume of the universe. Penrose and
Hawking \cite{44,45} have argued that expansion of the universe
has been started from this dense and hot phase by an explosion and
then universe is going to expand till today. After infinite time,
universe would have infinite volume with negligible density. In
our case, the proper energy density, string tension density as
well as particle density remain positive for increasing BD
parameter values and cosmic time. At initial epoch, the proper
energy and particle densities diverge while they turn out to be
finite for later time due to bulk viscous effects. Hence the
presence of viscosity prevents the universe to be empty in its
future evolution. Clearly, the physical behavior of our
constructed model supports these arguments as energy densities
turn out to be infinite (divergent) at initial epoch and hence is
of considerable interest. However, the string tension density
increases with the increase in time and BD parameter values.
\item Different parameters like $H,~H_{x},H_{y},~\Theta$ and $\sigma$ exhibit increasing behavior
for decreasing $\tau$ and become divergent at initial epoch
indicating the big bang start of the model. However, these
parameters approach to zero asymptotically, i.e.,
$\tau\rightarrow\infty$. Thus the constructed model has an initial
singularity. Moreover, scalar field evolves from zero (initially) to
infinite value as $\tau\rightarrow\infty$.
\item As we have discussed in our previous work \cite{24}, in case of BI
universe filled with perfect fluid, there is a constant and positive
deceleration parameter yielding decelerated expanding universe
model. This problem is then resolved by taking anisotropic matter
contents. In present work, the deceleration parameter turns out to
be a dynamical quantity rather than a constant due to the presence
of self-interacting potential, viscosity and magnetic field effects
which can provide negative values and hence yields accelerated
expansion of universe. For example, if we take
$\tau\rightarrow\infty$ and $-2<\omega<-3/2$ as well as decreasing
values of scalar field, the deceleration parameter gives negative
values lying in the range $-1<q<0$ which is in good agreement with
the observed range for cosmic expansion as shown in Figure
\textbf{2(b)}.
\item The anisotropic measure of expansion of the universe model turns out
to be a constant showing that the universe model exhibits
anisotropic behavior through the whole range of cosmic time. In
\cite{46}, it has been pointed out that some large-angle anomalies
are seen in CMB radiations, violating the statistical isotropy of
the observable universe. For a better description of these
anomalies, plane symmetric and homogeneous but anisotropic
universe models play a very significant role. Moreover, it is
found \cite{47}-\cite{49} that removing a Bianchi component in
WMAP data can explain various large-angle anomalies yielding an
isotropic universe. Thus the universe may have accomplished a
slight anisotropic geometry in cosmological models irrespective of
inflation. If the anisotropic nature of the model is maintained
through the whole range of time, then it may explain or discuss
these large-angle anomalies in CMB radiations.
\item The viscosity parameter increases with the passage of time
which corresponds to a non-empty universe in future.
\item In all physical parameters, the component of magnetic field has a
negative contribution, i.e., these physical parameters reduce due to
the presence of magnetic field.
\end{itemize}
In \cite{50}, a class of Bianchi II, VII and IX filled with cosmic
string fluid within Saez and Ballester gravity is discussed. It is
found that the constructed model has no initial singularity which
is inconsistent with the big bang model. Likewise, in a new work
\cite{51}, the same class is discussed with cosmological strings
in BD gravity and it is observed that the volume of the universe
model is contracting rather than expanding which is physically
unacceptable. However, in the present work, no such ambiguity
exists and the results are in well agreement with the
observations. It is worthwhile to mention here that our results
are consistent with those already available in the context of GR
\cite{27}.


\renewcommand{\theequation}{A\arabic{equation}}
\setcounter{equation}{0}

\end{document}